\documentclass[twocolumn]{aastex62}

\newcommand{\sub}[1]{_{\mbox{\scriptsize {#1}}}}

\accepted{Feb. 17th, 2020}

\shorttitle{FINAL MASSES OF GIANT PLANETS III}
\shortauthors{Tanaka et al.}

\begin{document}

\title{
FINAL MASSES OF GIANT PLANETS III:
EFFECT OF PHOTOEVAPORATION AND A NEW PLANETARY MIGRATION MODEL
}

\correspondingauthor{Hidekazu Tanaka}
\email{hidekazu@astr.tohoku.ac.jp}

\author{Hidekazu Tanaka}
\affil{Astronomical Institute, Tohoku University, Sendai 980-8578, Japan}

\author{Kiyoka Murase}
\affiliation{
Dept.~of Polar Science, School of Multidisciplinary Sciences, 
Graduate University for Advanced Studies, Tachikawa, 190-8518, Japan
}

\author{Takayuki Tanigawa}
\affiliation{
National Institute of Technology, Ichinoseki College, 
Ichinoseki 021-8511, Japan
}

\begin{abstract}
We herein develop a new simple model for giant planet formation, 
which predicts the final mass of a giant planet
born in a given disk, by adding the disk mass loss 
due to photoevaporation and a new type II migration formula 
to our previous model. 
The proposed model provides some interesting results. 
First, it gives universal evolution tracks in 
the diagram of planetary mass and orbital radius, which clarifies 
how giant planets migrate at growth
in the runaway gas accretion stage. 
Giant planets with a few Jupiter masses or less suffer only a slight 
radial migration 
in the runaway gas accretion stage.
Second, the final mass of giant planets is approximately 
given as a function of only three parameters: the initial 
disk mass at the starting time of 
runaway gas accretion onto the planet, 
the mass loss rate due to photoevaporation, and the starting 
time. On the other hand, the final planet mass is 
almost independent of the disk radius,
viscosity, and planetary orbital radius. The obtained final 
planet mass is $\lesssim$10\% of the initial disk mass. 
Third, the proposed model successfully explains properties in the 
mass distribution of giant exoplanets with the mass 
distribution of observed protoplanetary disks for a reasonable 
range of the mass loss rate due to photoevaporation.
\end{abstract}


\section{INTRODUCTION}
\label{sec:intro}
One of the main goals of the planet formation theory is to 
explain the statistics of thousands of exoplanets using the statistical 
properties of observed protoplanetary disks (e.g., Andrews et 
al.~2010) in a consistent manner. In the present paper, we 
try to explain the masses of giant exoplanets.

In the core accretion model, giant planets are formed with the 
following two growth stages in a protoplanetary disk (e.g., Mizuno 
1980). The first stage is the formation of the planetary solid core via
accumulation of planetesimals and pebbles (e.g., Kokubo \& Ida 2002; 
Lambrechts \& Johansen 2014). Once the solid core grow to the critical 
mass ($\sim$10$M_\oplus$) for gravitational collapse of its envelope, a 
rapid runaway accretion of the disk gas onto the planet proceeds after
a relatively slow contraction of the envelope (e.g., Pollack et al.~1996; 
Ikoma et al.~2000; Hubickyj et al.~2005), which is the second 
stage. In the present paper, we focus on the latter runaway gas 
accretion stage of the giant planet formation since the final masses of
the gaseous giant planets is governed by the runaway gas accretion. 
Although a fast type-I planetary migration at the stage of the solid 
core growth is also one of serious problems in giant planet formation 
(e.g., Tanaka et al.~2002; Paardekooper et al.~2011), this first stage 
is out of scope of the present study. 

The formation of giant planets is
governed by the gas accretion rate 
onto the planets and the radial migration speed. 
A massive planet with $\gtrsim$ several tens of earth masses opens a 
low-density gap around its orbit in the protoplanetary disk with the 
planetary torque (e.g., Lin \& Papaloizou 1986; Crida et al.~2006). The 
disk gap considerably changes the planetary growth and migration, as 
originally proposed by Lin \& Papaloizou (1986). Their original gap 
model assumes that almost no gas exists inside the gap and that no mass 
flow across the gap exists. Then the gap opening terminates growth of 
the planet. Because of no gap-crossing flow, the planet is locked in 
the gap and migrates with the disk viscous evolution, which is the
mechanism of the original type II migration. Many studies on planet 
formation have adopted the type II planetary migration and termination 
of growth due to the gap. 

However, two- and three-dimensional hydro-dynamical simulations showed 
that disk gas can easily cross the gap and also accrete onto the planet 
even for the gap created by a Jupiter-mass planet or larger (e.g., Kley 
1999; Lubow et al.~1999; Masset \& Snellgrove 2001; D'Angelo et al.~2003; 
Machida et al.~2010; Zhu et al.~2011). Thus the planetary growth 
continues even after the gap opening though the low-density gap reduces 
the mass accretion rate. The surface density inside the gap was examined 
by recent hydro-dynamical simulations and an empirical model of the gap 
surface density is proposed (e.g., Duffell \& MacFadyen 2013; Fung et al.~2014; 
Kanagawa et al.~2015a, 2016, 2017). Tanigawa \& 
Tanaka (2016, hereinafter Paper 2) constructed an analytic formula 
for the mass accretion rate onto the planet,
by using the empirical model for the 
gap surface density.
This formula reproduces 
very well the results of hydro-dynamical simulations by D'Angelo et 
al.~(2003) and Machida et al.~(2010). The present paper also uses 
this formula. 

The type II migration of giant planets has been 
problematic in previous studies on giant planet formation 
(e.g., Ida \& Lin 2004; Mordasini et al.~2009; Hasegawa \& Ida 2013;
Ida et al.~2018; Bitsch et al.~2015, 2019).
Duffell et al.~(2014) found a further faster type II migration
in their hydro-dynamical simulations.
Their obtained migration speeds are higher than 
those of the planet-dominated type II migration based on the 
original model (e.g., Armitage 2007)
by a factor of 3.
Their type II migration speeds\footnote{The migration 
of a gap-opening planet is still referred to as 
the type II migration though the assumptions in the original model 
was found to be invalid.
}
were confirmed by D\"urmann \& Kley (2015, 2017) and Robert et al.~(2018).
In Paper 2, however, we found that the reduction in the disk surface density
due to gas accretion onto the planet slows down the type II migration
sufficiently. Moreover, Kanagawa et al.~(2018) recently proposed a new model 
for type II migration. 
In their new model, the planet mainly interact 
with the disk gas inside the gap and the planetary migration speed
is proportional to the reduced surface density in the gap.
A similar idea on the type II migration had been first pointed out   
by D'Angelo and Lubow(2008).
The new model proposed by Kanagawa et al.\ reproduces very well the results 
obtained from the previous hydro-dynamical simulations of type II 
migration (Duffell et al.~2014; D\"urmann \& Kley 2015) and from 
the simulations by Kanagawa et al. The present paper uses 
their new formula for type II migration to revise our model.

In Paper 2, we also suggested that the final mass of a giant planet should 
be much more massive in the MMSN disk than the Jupiter mass. However, we 
did not take into account the disk mass loss due to the 
photoevaporation directly. Such a dissipation effect of photoevaporation 
is required in order to explain the disk lifetime and the 
rareness of the transitional disks (e.g., Clarke et al.~2001; 
Alexander et al.~2014). The present paper considers the disk mass 
loss due to photoevaporation, which would reduce the final planet
mass for a given disk.

Previous studies on the population synthesis of giant planets
have already examined the origin of statistics of exoplanets
in detail (e.g., Benz et al.~2004 and papers therein; Ida et al.~2018).
In order to examine the distributions of the final masses and orbital radii
of planets, they performed Monte Carlo simulations with probability 
distribution functions of various disk parameters. However, it is not clear 
yet what mass and orbital radius a planet will finally have 
in a given disk, because of uncertainties in disk models, 
the viscosity parameter, and models of planetary growth and migration.
Moreover, the dependence of the final planet mass on the disk parameters
(e.g., disk mass, radius, viscosity, and mass loss rate
due to photoevaporation) is unclear.
In addition, accurate empirical formulas for planetary growth rate
and migration speed mentioned above were not used properly in their
population synthesis calculations.

In the present paper, we revise our simple model for giant planet 
formation by including a new type II migration formula
and the disk mass loss due to photoevaporation by EUV.
We show that our new model can clearly predict the final mass 
of giant planets for a given disk, despite of many unfixed 
parameters in current disk models.

In Section 2, we briefly describe formulas of accretion rate and 
migration speed that have already been tested through hydro-dynamical 
simulations.
These formulas directly give universal evolution tracks 
in the diagram of planetary mass and orbital radius
for the runaway gas accretion stage,
although the prediction of the final planet mass also requires the disk 
model. In Section~3, we present a very simple disk model that includes
the disk mass loss due to photoevaporation by EUV.
Using this simple disk model, we derive a direct 
expression for the planetary growth rate.
In Section~4, we examine the time evolution and the final mass of 
giant planets for a reasonable parameter range. Our results for the 
final planet mass will also be applied to the origin of exoplanets.
In Section~5, we summarize our results.

\section{Models of growth and migration for giant planets} \label{sec:2}
\subsection{Assumptions} \label{subsec:21}
For giant planet formation, we adopt the core-accretion model.
We focus on the stage in which the solid core of a planet is more massive
than the critical core mass (Mizuno 1980, Ikoma et al.~2000).  
Then, the planet grows primarily via runaway accretion of the disk gas.
In the proposed model, as an initial condition, we assume that 
such a massive solid core exists in the disk.
For simplicity, we examine the growth and migration of a single giant planet in the 
gaseous disk and neglect the effect of other planets.

\subsection{Growth Rate and Migration Rate of a Planet} \label{subsec:22}

We adopt the rate of gas accretion onto a planet modeled by Paper 2.
The growth rate of the planet via the runaway gas accretion 
(or the accretion rate to the planet), $dM_p/dt$, is given by
\begin{equation}
\frac{d M_p}{dt} = D \, \Sigma\sub{gap}.
\label{dmpdt}
\end{equation}
%
The coefficient $D$ is empirically obtained as 
(Tanigawa \& Watanabe 2002)
\begin{equation}
D = 0.29 \left( \frac{M_p}{M_*} \right)^{4/3} 
         \left( \frac{h_p}{r_p} \right)^{-2}
         r_p^2 \Omega_p,
\label{D}
\end{equation}
%
where $M_*$ and $r_p$ are the mass of the central star and
the orbital radius of the planet, respectively.
The subscript $p$ of the scale height $h$ and the Keplerian angular 
velocity $\Omega$ indicates values at $r=r_p$.
The surface density in the planetary gap, $\Sigma\sub{gap}$,
is also given by an empirical formula (e.g., Duffell \& MacFadyen
2013; Kanagawa et al.~2015a, 2015b)\footnote{According to 
Kanagawa et al.~(2018), we use the prefactor 0.04 for $K$,
rather than 0.034 as was used in Paper 2.}
\begin{equation}
\Sigma\sub{gap} = \frac{\Sigma\sub{out}}{1+0.04 K},
\label{sigma_gap}
\end{equation}
%
where the non-dimensional parameter $K$ is given by
\begin{equation}
K = \left( \frac{M_p}{M_*} \right)^{2} 
    \left( \frac{h_p}{r_p} \right)^{-5} \alpha^{-1}.
\label{K}
\end{equation}
%
Moreover, $\Sigma\sub{out}$ is the surface density just outside of the gap,
and $\alpha$ is the viscosity parameter of the disk (Shakura \& Sunyaev 1973).
Note that Equation (\ref{sigma_gap}) derived from hydro-dynamical simulations
gives a much shallower gap than the previous one-dimensional analytic models 
(e.g., Lubow \& D'Angelo 2006; Tanigawa \& Ikoma (Paper 1)).
Owing to the shallow gap, the gas accretion onto the giant planet 
does not terminate even after the gap is formed.
The mass accretion rate of Equation~(\ref{dmpdt}) reproduces very 
well the results of hydro-dynamical simulations by D'Angelo et al.~(2003) 
and Machida et al.~(2010) for planets heavier than 10$M_\oplus$,
as shown in Figure~1 of Paper~2\footnote{Ginzburg \& Chiang (2019)
pointed out that, for planets less massive than 10$M_\oplus$,
the simulation results are well described by the
three-dimensional Bondi accretion rate rather than Equation~(\ref{dmpdt}).
However, we are interested in planets that have masses greater than 
the critical core mass ($\sim$10$M_\oplus$).
Equation~(\ref{dmpdt}) is better for those planets.}.

Exactly speaking, the gap surface density of Equation (\ref{sigma_gap}) 
represents the surface density around $r=r_p \pm 2 R\sub{Hill}$ rather than the 
lowest value at $r_p$ for a deep gap, where $R\sub{Hill}$ is the Hill radius 
of the planet (e.g., Fung et al.~2014; Kanagawa et al.~2017).
It would be reasonable that Equation (\ref{sigma_gap}) is used for the 
estimates of the accretion rate and the type II migration speed 
since those rates are determined by the surface densities around 
$r=r_p \pm 2 R\sub{Hill}$ rather than that at $r_p$
(Tanigawa \& Watanabe 2002; D'Angelo \& Lubow 2008).

It should be also noted that a slow Kelvin-Helmholtz contraction of 
the planetary envelope occurs between the stages of the solid core growth 
and the runaway gas accretion (e.g., Pollack et al.~1996; Ikoma et al.~2000).
The slow contraction of the envelope regulates the mass accretion rate 
for planets with $M_p \sim M\sub{crit}$. In this slow contraction stage, 
Equation~(\ref{dmpdt}) overestimates the mass accretion rate.
The timing of the transition from the slow contraction stage to the
runaway accretion stage where the accretion rate of Equation~(\ref{dmpdt})
is valid can be estimated as follows.
We can estimate the Kelvin-Helmholtz contraction time for the 
envelope of a planet with the mass $M_p$ as 
\begin{equation}
\tau\sub{KH} \simeq 1 \times 10^3
\left( \frac{M_p}{30M_\oplus} \right)^{-2.5}
\left( \frac{\kappa}{0.05\mbox{cm}^2 \mbox{g}^{-1}} \right)
\mbox{yr},
\label{taukh}
\end{equation}
%
by replacing the core mass $M\sub{core}$ with $M_p$ in the analytic 
contraction time derived by Ikoma et al.~(2000). 
In Equation~(\ref{taukh}), $\kappa$ is the opacity of the planetary 
envelope. The opacity of $\kappa$ =0.05 g$\,$cm$^{-2}$ is consistent
with Movshovitz \& Podolak (2008) who calculated the envelop opacity,
taking into account dust growth and sedimentation in the envelope.
A contraction time similar to Equation~(\ref{taukh}) is adopted in the 
planetary population synthesis calculations (e.g., Ida \& Lin 2004, 
Benz et al.~2014, Ida et al.~2018). 
Equation~(\ref{taukh}) can also explain the mass accretion rates obtained by 
Lambrechts et al.~(2019) with their three-dimensional non-isothermal 
hydro-dynamical simulations for $M_p < 100M_\oplus$,
by using the value of the opacity they adopted. 
The mass accretion rate regulated by the slow contraction is given 
by $M_p/\tau\sub{KH}$.
By equating this accretion rate with Equation~(\ref{dmpdt}),
we can estimate the planet mass at the transition from the slow contraction 
to the runaway accretion
as $30 (\kappa/0.05 \mbox{cm}^2 \mbox{g}^{-1})^{6/13} M_\oplus$
for $\Sigma\sub{gas}=\mbox{100g}\,\mbox{cm}^2$ and the disk aspect ratio of
$h/r=0.05$ at 5AU.
Hence, for planets with $M_p>30 M_\oplus$, the envelope contraction
is rapid enough and Equation~(\ref{dmpdt}) is valid.

The radial migration speed of the planet is generally expressed with the 
torque on a planet, $\Gamma$, as
\begin{equation}
\frac{d}{dt} \ln r_p 
= 2 \frac{d}{dt} \ln L_p 
= \frac{2 \Gamma}{ M_p r_p^2 \Omega_p},
\label{drdt0}
\end{equation}
%
where $L_p$ is the angular momentum of the planet.
Kanagawa et al.~(2018) found that by using the gap surface density
$\Sigma\sub{gap}$, the torque on a planet embedded in the disk gap
(i.e., the torque of type II planetary migration) is given by an expression 
similar to that for the type I torque (e.g., Tanaka et al.~2002):
\begin{equation}
\Gamma = -3.0 \left( \frac{M_p}{M_*} \right)^{2} 
               \left( \frac{h_p}{r_p} \right)^{-2}
               r_p^4 \Omega_p^2 \Sigma\sub{gap}.
\label{gamma}
\end{equation}
%
%
From Equations (\ref{drdt0}) and (\ref{gamma}), the migration speed of 
a planet is given by
\begin{equation}
\frac{d}{dt} \ln r_p = 
-6.0 \, \frac{M_p}{M_*} 
\left( \frac{h_p}{r_p} \right)^{-2}
\frac{r_p^2\, \Sigma\sub{gap}}{M_*}
\Omega_p.
\label{drdt}
\end{equation}
%

The migration speed for type II migration also reproduces very well
the results obtained from the previous hydro-dynamical simulations of
type II migration (D\"urmann \& Kley 2014, Duffell et al.~2014)
and from the simulations by Kanagawa et al.
Type II migration was previously thought to be caused by the
interaction with gap edges (i.e., outside of the gap) (e.g., Lin \&
Papaloizou 1986, 1993; Armitage 2007). 
On the other hand, the new more accurate formula by
Kanagawa et al.~indicates that the planet mainly interacts with
the gas inside the gap. 
For a less massive planet forming no gap, Kanagawa et al's formula
is reduced to that of type I migration.
In the present paper, we use this new formula for the type II
planetary migration speed, although Paper 2 basically adopted the 
previous type II migration model.

The present paper does not include type III migration or dynamical 
corotation torque which appear only in very massive disks
(e.g., Masset \& Papaloizou 2003; Paardekooper 2014; Pierens \& Raymond 2016).
Pierens \& Raymond (2016) showed with their hydro-dynamical simulations
that planets experience a more rapid (outward) migration than type II 
due to the dynamical corotation torque during their growth.
In their simulations, however,
planets grow to about 10 jupiter masses within $1 \times 10^4$ yrs. 
Thus it is expected that those bodies might grow to several tens of 
jupiter masses (i.e., the mass range of brown dwarfs or red dwarfs)
within the typical disk lifetime of $\sim 10^6$ yrs.
Formation of such massive bodies are beyond the scope of the present study.

\subsection{Universal Evolution Tracks in the Mass-Orbit 
Diagram} \label{subsec:23}

In our model, the growth rate and the migration speed are both
proportional to $\Sigma\sub{gap}$. The time evolution of the planet
mass and orbital radius depends on $\Sigma\sub{gap}(t)$ or the
disk model described in the next section.
Here, we consider planetary evolution tracks in the diagram of mass and orbital
radius. Interestingly, the evolution tracks are independent of the model of
the protoplanetary disk, as shown below.

Dividing Equation~(\ref{dmpdt}) by (\ref{drdt}) and using (\ref{D}), 
we obtain a simple differential equation for the evolution tracks:
\begin{equation}
\frac{d \ln M_p}{d \ln r_p} 
= \frac{0.29}{6.0} \left( \frac{M_p}{M_*} \right)^{-2/3}
= \left( \frac{M_p}{M\sub{th}} \right)^{-2/3},
\label{dlnmdlnr}
\end{equation}
%
where the 
threshold mass $M\sub{th}$ is given by
\begin{equation}
M\sub{th} = \left( \frac{0.29}{6.0} \right)^{3/2} M_*
= 0.011 M_*.
\label{mth}
\end{equation}
%
This simple form of Equation~(\ref{dlnmdlnr})
is available because $d M_p/dt$ and $d \ln r_p/dt$ are both proportional to
$(h_p/r_p)^{-2} \Sigma\sub{gap}$.
Solving Equation~(\ref{dlnmdlnr}), we obtain an analytic expression 
for the universal evolution tracks in the diagram of planetary mass 
and orbital radius as
\begin{equation}
\frac{r_p}{r_0} = 
\exp \left \{ -\frac{3}{2} 
\left [ 
\left( \frac{M_p}{M\sub{th}} \right)^{2/3}
- \left( \frac{M_0}{M\sub{th}} \right)^{2/3}
\right ]
\right \},
\label{evo_curve}
\end{equation}
%
where $M_0$ and $r_0$ are the initial mass and initial orbital radius
of the planet, respectively. We recall that the obtained evolution 
tracks are completely independent of the disk model.
Our evolution tracks are independent of the disk gap model of 
Equation~(\ref{sigma_gap}), too.

\begin{figure}[ht]
\begin{center}
%
\includegraphics[width=8.3cm]{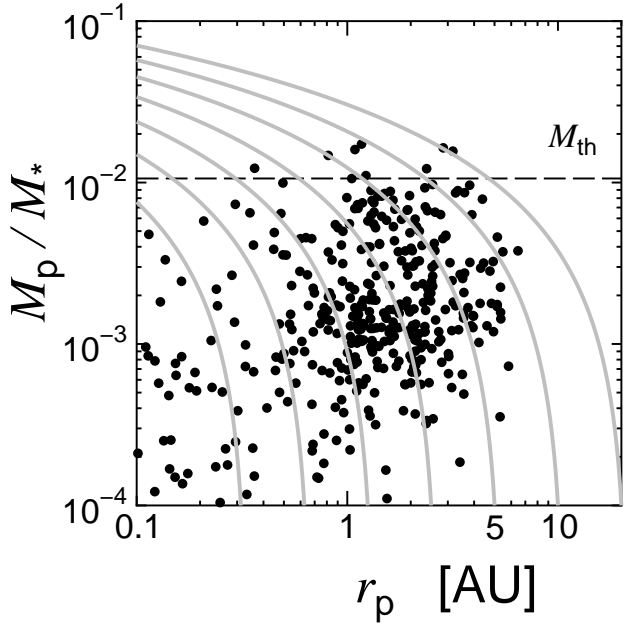}
%
\caption{Universal evolution tracks of giant planets in the mass-orbit diagram  
in the runaway gas accretion stage. Our evolution tracks are completely 
independent of the disk model whereas the disk model affects termination of 
planetary growth on each track.
{The initial mass, $M_0$, is set to be $10^{-4} M_*$
and the initial orbital radii, $r_0$, are varied from 0.31AU to 20AU.}
Data of exoplanets observed by the radial-velocity method 
(http://exoplanets.org) are also plotted.
As for masses of exoplanets, the observed values of $M_p \sin i$ are simply used.
}
\label{fig:1}
\end{center}
\end{figure}

Figure~\ref{fig:1} shows the universal evolution tracks of 
Equation~(\ref{evo_curve}) in the diagram of planet mass and orbital 
radius. In Figure~\ref{fig:1}, we also plot the data of exoplanets 
observed by the radial-velocity method, 
which have a weak observational bias compared with the radial 
distribution of the exoplanets discovered by transit surveys.
Planets less massive than the threshold mass $M\sub{th}$
($\simeq \! \! 10M_J$ for a solar-mass star) do not suffer much radial migration. 
This is because the migration timescale is longer than the growth 
time for $M_p < M\sub{th}$.
The final orbital radius of a planet with 
$M\sub{th}$ is approximately 1/5 of the initial orbital radius. Previous 
studies reported a problematic rapid type II migration in the giant 
planet formation, as described in Introduction. 
However, our model shows that very massive exoplanets 
exceeding $M\sub{th}$ plotted in Figure~\ref{fig:1} can also be 
formed from solid cores initially located within 20 AU. 
Our model succeeds in fixing the problem of type II migration
in giant planet formation\footnote{The evolution tracks derived by 
Ida et al.~(2018) depend on the viscosity parameter
although they adopted the gas accretion rate of Paper 2 and the 
migration formula by Kanagawa et al.~(2018) as well as our model.
By introducing two kinds of viscosity parameters, they managed to avoid 
rapid type II planetary migration.
In Section~\ref{subsec:34}, we will explain the difference in migration 
prescription between Ida et al.\ and the present paper.}.

Slight migrations for Jupiter-mass planets or less, on the other hand,
make the origin of hot Jupiters difficult to explain.
Orbital radii of planets can be shifted by mutual interactions between 
multiple planets, which are not included in our model.
As an explanation of hot Jupiters, the model that considers the planet-planet 
scattering followed by the tidal circularization 
(e.g., Rasio \& Ford 1996; Nagasawa et al.~2008; Winn et al.~2010)
is more plausible than the model that considers type II migration.
The type I migration of solid planetary cores before the runaway gas 
accretion stage can be an alternative possible origin of hot Jupiters.
Solid planetary cores can easily migrate inward
since their growth time is expected to be much longer than
that at the runaway gas accretion stage.

A relatively large number of giant exoplanets are observed in the 
radial range of from 1 to 3 AU in Figure~\ref{fig:1}. These crowded 
exoplanets can be explained if a large number of massive solid cores 
are formed from 1.5 AU to 4 AU of protoplanetary disks. 
Such massive planetary embryos may be naturally formed just 
outside of the snow line, which is located at 1 to 3 AU.

Note that the disk model affects where and how the planetary growth 
terminates on an evolution track (i.e., the final mass and location 
of the planet). We will discuss these items, using a simple disk 
model described in the next section.

It should be also noticed that the torque and the migration speed in 
Kanagawa et al's formulas (Equations (\ref{gamma}) and (\ref{drdt})) can 
be underestimated in the absolute values in the regime of type I migration 
(i.e., for $M_p \lesssim 10^{-4} M_*$). Although the coefficient of the 
torque is fixed as $-3.0$ in Equation (\ref{gamma}), the torque depends on 
the radial gradients of the disk surface density and temperature in the 
accurate type I formulas (e.g., Tanaka et al.\ 2002; Paardekooper 
et al.\ 2011). Owing to steep radial distributions in $\Sigma$ and/or $T$,
the accurate type I migration can be faster than Equation~(\ref{drdt}) 
by the factor of 2 or 3.
However, such a possible enhancement in the migration speed would hardly 
change the result of a slight migration for $M_p \sim 10^{-4}M_*$ because 
the migration timescale given by Equation~(\ref{drdt}) is about 20 times 
as long as the growth time at $M_p \sim 10^{-4}M_*$.
\section{Disk model} \label{sec:3}
\subsection{Self-similar Solution of Accretion Disks} \label{subsec:31}
Our simple disk model is based on the self-similar solution of
accretion disks with the viscosity $\nu$ being proportional to $r^\gamma$
(Lynden-Bell \& Pringle 1974; Hartmann et al.~1998).
The temperature distribution can be more complex than a single power-law 
in realistic disk models. Such a complex disk model should be examined in the 
future work.
In this simple model, we also include the effects of photoevaporation of disk gas 
by EUV radiation from the central star
and gas accretion onto a planet. We assume that the orbital radius 
of the planet $r_p$ and the radial location $r_g$, where 
the photoevaporation primarily occurs, are much smaller than the disk radius.
Then, the planet and the photoevaporation would have only minor effects
on the evolution of the outer part of the disk, which contains most of the 
(total) disk mass.
Gorti \& Hollenbach (2009) suggested that FUV radiation from the central 
star can cause rapid erosion of the outer disk. Such strong photoevaporation 
by FUV can reduce the disk mass in a relatively short time ($\sim 10^6$yr). 
However, since the mass loss rate by FUV photoevaporation is still uncertain 
to about an order of magnitude (e.g., Alexander et al.~2014), we do not include 
photoevaporation by FUV in our disk model.

At an outer part of the disk, the surface density is then simply given by
the self-similar solution:
\begin{equation}
\Sigma (r,t) 
= \frac{M_d(t)}{6 \pi (2-\gamma) \nu t}
\exp\left[ - \left( \frac{r}{R_d(t)} \right)^{2-\gamma} \right].
\label{sigmass}
\end{equation}
%
where $\nu=\nu_0 r^\gamma$ and 
the characteristic disk radius $R_d$ is given by
\begin{equation}
R_d = [\, 3(2-\gamma)^2 \nu_0 \, t \,]^{\frac{1}{2-\gamma}}.
\label{rdss}
\end{equation}
%
The disk mass $M_d$ decreased as follows:
\begin{equation}
M_d(t) = M_d(t_0) \left( \frac{t}{t_0} \right)^{-\frac{1}{4-2\gamma}}.
\label{mdss}
\end{equation}
%

At the intermediate disk region in which $r_p, r_g \ll r \ll R_d$,
the surface density and the disk accretion rate $\dot{M}_d$ are 
approximately given by
\begin{equation}
\Sigma = \frac{M_d(t)}{3 \pi (4-2\gamma) \nu t}, \qquad \! \!
\dot{M}_d = \frac{M_d(t)}{(4-2\gamma) t}  \!\!
\qquad (\mbox{for } \: \: r \ll R_d)
\label{sigmamddotss}
\end{equation}
%
%
and we obtain the well-known relation between them as
\begin{equation}
\Sigma = \frac{\dot{M}_d}{3 \pi \nu} 
\qquad \qquad(\mbox{for } \: \: r \ll R_d).
\label{sigma2ss}
\end{equation}
%
Note that the time-evolution of the disk mass 
and the accretion rate are independent of
the disk viscosity in Equations (\ref{mdss}) and (\ref{sigmamddotss}).
Then, the disk surface density is inversely proportional to the viscosity.

We assume the disk temperature as $280 (r/1\mbox{AU})^{-1/2} $K
(Hayashi et al.~1985).
Then, the disk aspect ratio (i.e., the ratio of the scale height 
to the radius) is given by
\begin{equation}
\frac{h}{r} = 0.05 \left(\frac{r}{5\mbox{AU}} \right)^{1/4}.
\label{scaleheight}
\end{equation}
Using the parameter $\alpha$,
the disk viscosity is also expressed as 
\begin{equation}
\nu = \alpha h^2 \Omega.
\label{viscosity}
\end{equation}
In the next section, we adopt $\gamma =1$ in the nominal case,
assuming a constant $\alpha$.

\subsection{Effect of Photoevaporation} \label{subsec:32}
In our model, we regard the mass loss rate due to the photoevaporation
$\dot{M}\sub{w}$ as a parameter. We assume that the mass loss due to 
photoevaporation occurs primarily outside the planet orbit.
Even for the case in which the photoevaporation inside the planet orbit
is not negligible, the following treatment for photoevaporation
would also be valid by considering $\dot{M}\sub{w}$ 
as the mass loss rate only outside of the planet orbit.

If no mass loss due to photoevaporation exists,
then the mass supply rate to the planet-forming inner region,
$\dot{M}\sub{sup}$, is equal to the disk accretion rate,
$\dot{M}_d$. When the photoevaporation is effective,
the mass supply rate to the planet-forming region is given by
\begin{equation}
\dot{M}\sub{sup} = \dot{M}_d - \dot{M}\sub{w}
\label{msupdot}
\end{equation}
%
%
if $\dot{M}_d \ge \dot{M}\sub{w}$. 
Then, the disk surface density in the planet-forming inner
region is given by
\begin{equation}
\Sigma = \frac{\dot{M}\sub{sup}}{3 \pi \nu} 
=  \frac{\dot{M}_d-\dot{M}\sub{w}}{3 \pi \nu}.
\label{sigma_photo}
\end{equation}
%

The disk accretion rate decreases gradually.
The time at which planet growth stops, $t\sub{end}$,
is determined by the equation $\dot{M}\sub{sup} =0$, i.e.,
\begin{equation}
\dot{M}_d(t\sub{end}) = \dot{M}\sub{w}.
\label{tend}
\end{equation}
%
Using Equation~(\ref{sigmamddotss}),
we can rewrite Equation~(\ref{tend}) as
\begin{equation}
\left( 
\frac{t\sub{end}}{t_0} 
\right)^{\frac{5-2\gamma}{4-2\gamma}} 
= 
\frac{{M}_d(t_0)}{(4-2\gamma) t_0 \dot{M}\sub{w}}.
\label{tend2}
\end{equation}
%
%
We regard $t_0$ as the starting time of the runaway gas accretion 
onto the planet, and $M_d(t_0)$ is the disk mass at the starting time.
When $M_d(t_0) < (4-2\gamma)t_0 \dot{M}\sub{w}$, 
Equation~(\ref{tend2}) gives $t\sub{end} < t_0$.
This means that runaway gas accretion onto the planet cannot 
occur because of gas dissipation at the planet-forming region before $t_0$
due to strong photoevaporation.
By integrating Equation~(\ref{msupdot}) from $t_0$ to $t\sub{end}$,
we can also obtain the total gaseous mass supplied 
to the planet-forming region from $t_0$ up to $t\sub{end}$,
$M\sub{sup}$,
as
\begin{eqnarray}
M\sub{sup} &=& \int^{t\sub{end}}_{t_0} \dot{M}\sub{sup} \, dt  
\nonumber \\
&=& {M}_d(t_0) 
\left[ 
1 - \left( \frac{t\sub{end}}{t_0} \right)^{-\frac1{4-2\gamma}} 
\right]  
- \dot{M}\sub{w} (t\sub{end}-t_0) \nonumber \\
&=& {M}_d(t_0)
\left[
1 - \frac{5-2\gamma}{4-2\gamma} 
\left( \frac{t\sub{end}}{t_0} \right)^{-\frac1{4-2\gamma}} 
\right]
+ \,
t_0 \dot{M}\sub{w}.
\nonumber \\
\label{msup}
\end{eqnarray}
%

%
\subsection{Effects of Gas Accretion onto the Planet} \label{subsec:33}
\begin{figure}[t]
\begin{center}
%
\includegraphics[width=8.5cm]{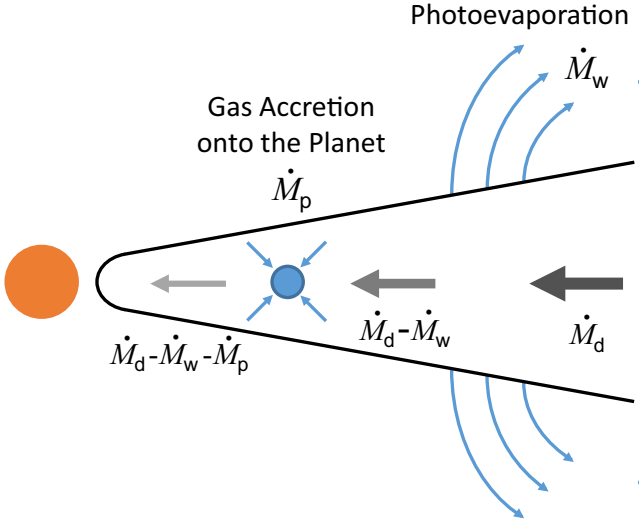}
\caption{Schematic diagram of the proposed simple disk model.
Because of photoevaporation with mass loss rate $\dot{M}\sub{w}$, 
the mass supply rate to the planet-forming region is reduced to 
$\dot{M}_d - \dot{M}\sub{w}$.
The disk accretion rate at the innermost disk is further 
reduced to $\dot{M}_d - \dot{M}\sub{w} - \dot{M}_p$
due to gas accretion onto the planet.
From these disk accretion rates, the surface density distribution 
of this viscous accretion disk is obtained in the proposed disk model.
}
%
\label{fig:12}
\end{center}
\end{figure}

Lubow \& D'Angelo (2006) gave the analytic expression of the surface 
density reduced by the gas accretion onto a planet 
as\footnote{The derivation of this analytic expression is shown in 
Appendix~B of Paper 2. Paper 2 also gives the radial surface density 
distribution.}:
\begin{equation}
\Sigma\sub{out}(r_p) 
= \frac{\dot{M}_d - \dot{M}_p}{3 \pi \nu(r_p)}.
\label{sigmaout2d}
\end{equation}
%
In the above, we used the notation, $\Sigma\sub{out}$, which is the
surface density just outside the planetary gap.
This is because the effect of the planetary gap is not included
in Equation~(\ref{sigmaout2d}).
This expression is not valid when the photoevaporation is effective.
However, we can readily include the effect of photoevaporation in
Equation~(\ref{sigmaout2d}), by simply replacing $\dot{M}_d$ with 
$\dot{M}\sub{sup}$:
\begin{equation}
\Sigma\sub{out}(r_p) 
= \frac{\dot{M}_d-\dot{M}\sub{w} - \dot{M}_p}{3 \pi \nu(r_p)}.
\label{sigmaout2}
\end{equation}
%
This is the expression of $\Sigma\sub{out}$ including both 
the effects of photoevaporation and gas accretion onto the planet.
The mass flux crossing the gap to the innermost disk
is given by $\dot{M}_d-\dot{M}\sub{w}-\dot{M}_p$
and it is reduced by the gas accretion onto the planet
and the photoevaporation.
In Figure~\ref{fig:12}, we summarize our simple disk model.

\subsection{Direct Expression for the Planetary Growth Rate} \label{subsec:34}
The growth rate, $\dot{M}_p$, included in Equation~(\ref{sigmaout2}) 
depends on $\Sigma\sub{gap}$ (or $\Sigma\sub{out}$), 
as shown in Equation~(\ref{dmpdt}).
By solving the coupled equations (\ref{dmpdt}), (\ref{sigma_gap}), 
and (\ref{sigmaout2}), we obtain
\begin{equation}
\Sigma\sub{gap} = 
\frac1{1+0.04K}
\left(1 + \frac{D'}{3\pi \nu(r_p)} \right)^{-1}
\frac{\dot{M}_d-\dot{M}\sub{w}}{3 \pi \nu(r_p)}
\label{sigmaout3}
\end{equation}
and
\begin{equation}
\dot{M}_p = \frac{D'/3\pi \nu(r_p)}{1+D'/3\pi \nu(r_p)} \,
(\dot{M}_d-\dot{M}\sub{w}),
\label{dmdt2}
\end{equation}
where $D'$ is given by
\begin{equation}
D'=\frac{D}{1+0.04K}
\label{ddash}
\end{equation}
and $D$ and $K$ both depend on $M_p$ and $r_p$,
as shown in Equations~(\ref{D}) and (\ref{K}).
The first and second factors in the RHS of Equation~(\ref{sigmaout3}) 
represent the reduction factors of the surface density due to
the planetary gap and due to the gas accretion onto the planet,
respectively. 
Both factors reduces the type II migration speed
as well as the gas accretion rate onto the planet
since these rates are proportional to $\Sigma\sub{gap}$.
%
%
Equation~(\ref{dmdt2}) gives a direct expression for the planetary
growth rate. Note that the disk accretion rate $\dot{M}_d$ is
given by Equation~(\ref{sigmamddotss}) and that the orbital radius 
$r_p$ is dependent on $M_p$, as show in Equation~(\ref{evo_curve}).

When the ratio $D'/(3\pi \nu)$ is large, 
the mass accretion rate $\dot{M}_p$ is
nearly equal to the supply rate $\dot{M}_d - \dot{M}\sub{w}$.
This means that an insufficient mass supply rate
(or a low disk accretion rate) regulates $\dot{M}_p$. The regulation of
$\dot{M}_p$ due to a small disk accretion rate is taken into account 
in many population synthesis calculations.
The regulation of $\dot{M}_p$ are originated from the surface density
reduction due to the second factor in Equation~(\ref{sigmaout3}).
This surface density reduction due to a low disk accretion rate
also slow down the type II migration, as first pointed out by
Tanigawa \& Tanaka (2016). 
Robert et al.~(2018) also found the slow-down of type II migration
due to a rapid gas accretion on the planet
in their hydro-dynamical simulations.
However, recent population synthesis calculations does not
include the slowing mechanism yet, even though they include the
regulation of $\dot{M}_p$ due to a low disk accretion rate (e.g.,
Ida et al.~2018, Johansen et al.~2019, Bitsch et al.~2019).
Such an inconsistent treatment in the growth rate and the migration speed
breaks down the universality of the evolution tracks proposed
in Section 2.3.

Estimating the non-dimensional ratio $D'/(3\pi \nu)$ is valuable.
We consider a deep-gap case in which $0.04 K \gg 1$.
From Equation~(\ref{K}), this corresponds to the case of relatively 
massive planets having masses that satisfy
\begin{equation}
M_p \gg 0.1 \left( \frac{h/r}{0.05} \right)^{5/2} 
\left( \frac{\alpha}{10^{-3}} \right)^{1/2} M_J.
\end{equation}
The ratio is then estimated as
\begin{eqnarray}
\frac{D'}{3\pi \nu} 
&\simeq& \frac{D/(0.04 K)}{3\pi \nu} 
= 0.8 \left( \frac{M_p}{M_*} \right)^{-2/3}
\frac{h_p}{r_p} \nonumber \\[1mm]
&=& 1 \left( \frac{M_p}{7 M_J} \right)^{-2/3}
\left( \frac{M_*}{M_\odot} \right)^{2/3}
\frac{h_p/r_p}{0.05}.\\
\nonumber
\label{ddash3pinu}
\end{eqnarray}
For a planet much less massive than $7M_J$ (at 5 AU around a solar-mass 
star), $D'/(3\pi \nu) \gg 1$, and Equation~(\ref{dmdt2}) gives
$\dot{M}_p = \dot{M}\sub{sup}$. That is, the gas supplied to the 
planet-forming region almost perfectly accretes onto the planet.
Almost perfect accretion onto the planet reduces much the mass flux 
across the gap. Hydro-dynamical simulations by Zhu et al.~(2011) also show 
similar reductions in the mass flux to the innermost disk by rapid gas 
accretion onto the planet.
For a planet more massive than 7$M_J$, on the other hand, the first factor 
in the RHS of Equation~(\ref{dmdt2}) becomes small. Then, only a minor portion 
accretes onto the planet, and most of the gas flows into the innermost disk 
in spite of a deep gap created by the massive planet.
Although one may expect that a deep gap might prevent the gas flow across the 
gap, hydrodynamical simulations show that disk gas can easily cross even a deep 
gap formed by a jupiter-mass planet (e.g., Zhu et al.~2011; 
D\"urmann \& Kley 2015, 2017).

Moreover, note that $D'/(3\pi \nu)$ is independent of the viscosity parameter
$\alpha$ when $0.04 K \gg 1$. Then, from Equations~(\ref{dmdt2})
and (\ref{dlnmdlnr}), we find that the time evolution rates $dM_p/dt$
and $dr_p/dt$ are also independent of $\alpha$.
Although there still exists a large uncertainty in the value of $\alpha$, 
we can discuss the growth and migration of a giant planet embedded in a 
protoplanetary disk independently of $\alpha$ using the proposed 
model\footnote{As seen from Equation~(\ref{sigmaout3}), the surface density 
in the gap, $\Sigma\sub{gap}$, is also independent of $\alpha$, 
for $0.04K \gg 1$. 
The ratio of $\Sigma\sub{gap}/\Sigma\sub{out}$ representing the gap depth
is proportional to $\nu$ while $\Sigma\sub{out} \propto \nu$ in 
Equation~(\ref{sigmaout2d}). Since these two viscosity-dependences cancel 
out, $\Sigma\sub{gap}$ and $\dot{M}_p$ are almost independent of $\nu$.
}
\section{results} \label{sec:4}
\subsection{Final Planet Masses in Various Disks} \label{subsec:41}
Under our simple disk model described in last section, 
we can calculate the time evolution of the planet mass,
by integrating Equation~(\ref{dmdt2}) with Equations
(\ref{D}), (\ref{K}), (\ref{evo_curve}), (\ref{mdss}),
(\ref{scaleheight}), (\ref{viscosity}), and (\ref{ddash}).
These equations have six parameters, i.e., the viscosity 
parameter $\alpha$, the mass loss rate due to photoevaporation 
$\dot{M}\sub{w}$, the starting time of the runaway gas accretion 
onto the planet $t_0$, the initial disk mass $M_d(t_0)$,
the initial planet mass $M_0$, and the initial orbital radius $r_0$
of the planet\footnote{The initial disk radius, $R_d(t_0)$,
is a function of the parameters $t_0$ and $\alpha$
(see Equations~(\ref{rdss}) and (\ref{viscosity})).
For the nominal values of $\alpha$ and $t_0$, $R_d(t_0)$ is 42 AU.}.
%
%
As nominal values, we set
$\alpha=10^{-3}$, $M_0=6\times 10^{-5}M_*$ ($=20M_\oplus$),
$r_0=5$ AU, and $t_0=2\times 10^6$ yr.
In our model, however, the final masses of giant planets depend on  
$\alpha$, $M_0$, and $r_0$ only weakly, as will be shown in 
Figure~\ref{fig:3}. In the calculation of the final planet mass,
$t_0$ and $\dot{M}\sub{w}$ are included only in the form of 
the product $t_0 \dot{M}\sub{w}$ (e.g., Equation (\ref{tend2}))
if we use the normalized time $t/t_0$. 
Thus, we can consider this product as a single parameter.
In the nominal case, we also set $\gamma =1$, 
assuming a constant $\alpha$ (Section~\ref{subsec:32}).
The disk observations also suggest $\gamma \simeq 1$ 
(e.g., Andrews et al.~2010).
The cases with $\gamma \ne 1$ will be shown in Figure~\ref{fig:8}.

\begin{figure}[t]
\begin{center}
%
\includegraphics[width=8.2cm]{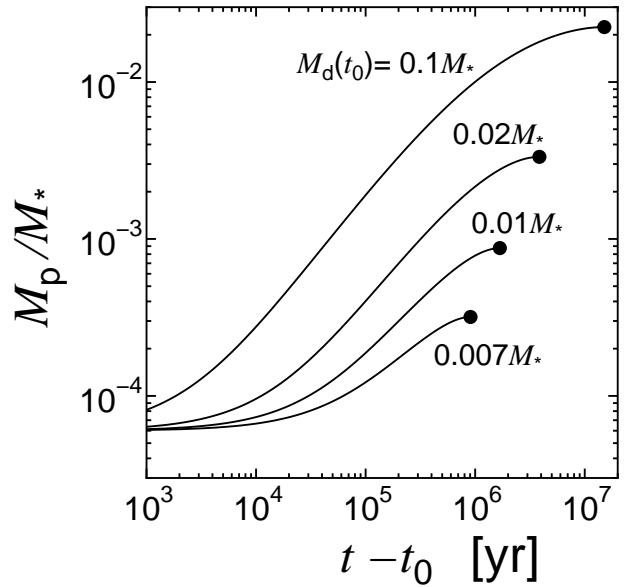}
%
\caption{
Time evolution of the planet mass in the runaway gas accretion stage
for various initial disk masses
in the case of $\dot{M}\sub{w}=10^{-9} M_* /\mbox{yr}$.
The initial disk masses $M_d(t_0)$ are 0.007, 0.01, 0.02, 
and $0.1 M_*$
in the case of $\dot{M}\sub{w}=10^{-9} M_* /\mbox{yr}$. 
Other parameters are set to be nominal values (see the text).
Filled circles represent the final planet mass at 
$t\sub{end}$.
}
\label{fig:2}
\end{center}
\end{figure}

Figure~\ref{fig:2} shows the time evolution of the planet mass
via the gas accretion for various initial disk masses $M_d(t_0)$
in the case of $\dot{M}\sub{w}=10^{-9} M_* /\mbox{yr}$.
Note that these planets radially migrate
along the evolution track starting from 5AU of Figure~\ref{fig:1}.
The gas accretion onto the planet terminates due to the 
photoevaporation at $t\sub{end}$. Filled circles represent
the final planet mass at $t\sub{end}$.
The final planet mass increases with the initial disk mass.
A Jupiter-mass planet is produced in a disk
with $M_d(t_0)=0.01 M_\odot$ for $M_*=M_\odot$ and 
$\dot{M}\sub{w}=10^{-9} M_\odot/\mbox{yr}$.

\begin{figure}[b]
\begin{center}
%
\includegraphics[width=8.2cm]{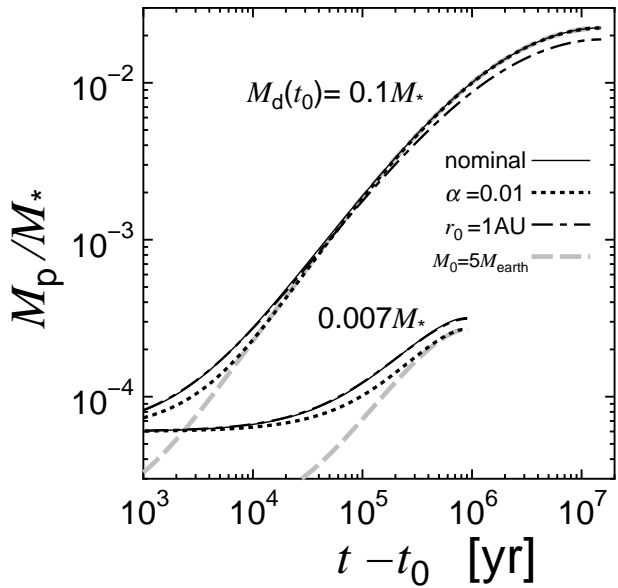}
%
\caption{Same as Fig.~\ref{fig:2}
but with one of the parameters $\alpha$, $M_0$, or $r_0$
varied from the nominal values for $M_d(t_0)= 0.007$ and $0.01 M_*$.
The solid lines represent nominal cases. For dotted lines, $\alpha$
is set to be 0.01. The initial planet mass and orbital radius
are five earth masses and 1 AU for gray dashed lines and dot-dashed lines,
respectively.
}
\label{fig:3}
\end{center}
\end{figure}
\begin{figure}[hb]
\begin{center}
%
\includegraphics[width=8.2cm]{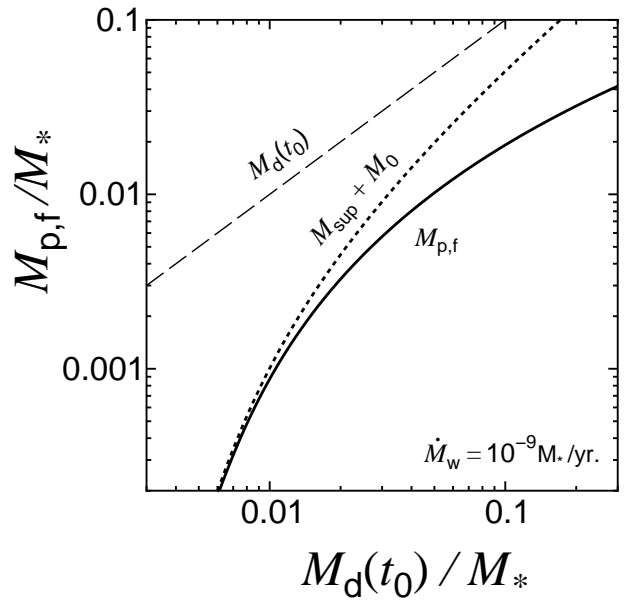}
%
\caption{Final planet mass as a function of the initial disk mass $M_d(t_0)$ 
(solid line) in the case of $\dot{M}\sub{w}=10^{-9} M_* /\mbox{yr}$. 
For comparison, $M_d(t_0)$ (dashed line) and $M\sub{sup}+M_0$ (dotted line) 
are also plotted. The latter is the planet mass when all of the gas supplied 
from the outer disk perfectly accretes onto the planet.
All masses are normalized by $M_*$.
}
\label{fig:4}
\end{center}
\end{figure}
\begin{figure}[ht]
\begin{center}
%
%
\includegraphics[width=8.4cm]{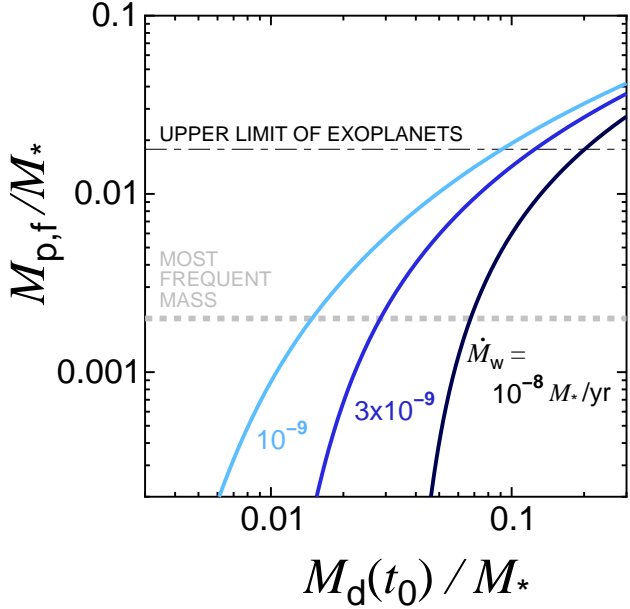}
%
\caption{Same as Figure~\ref{fig:4} but for various mass loss rates 
due to photoevaporation, $\dot{M}\sub{w}$. 
The light blue, blue, and black lines represent
the cases with $\dot{M}\sub{w}=10^{-9}$, $3\times10^{-9}$, and
$10^{-8} M_* /\mbox{yr}$, respectively.
The dot-dashed line is the upper mass limit of exoplanets, 
and the gray dotted line indicates the most frequent 
mass of exoplanets in Figure~\ref{fig:1}.}
\label{fig:5}
\end{center}
\end{figure}
\begin{figure}[ht]
\begin{center}
%
\includegraphics[width=8.2cm]{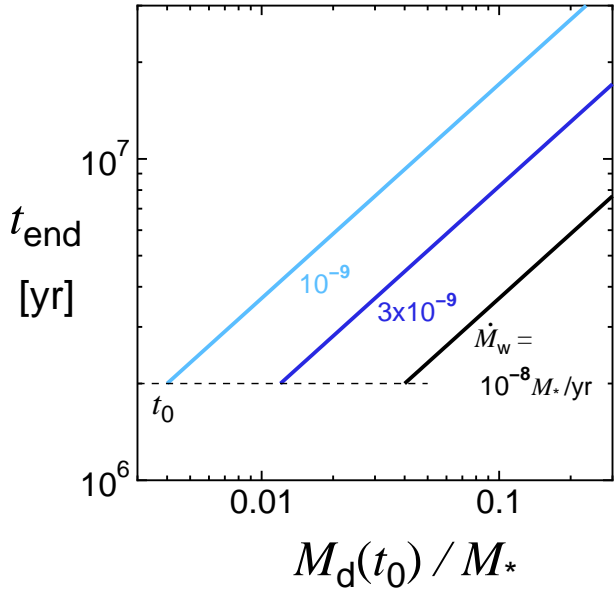}
%
\caption{Disk dissipation time, $t\sub{end}$,
at the planet-forming region
as function of the initial disk mass $M_d(t_0)$
for various mass loss rates.}
\label{fig:6}
\end{center}
\end{figure}

In Figure~\ref{fig:3}, we check for the dependence of the time evolution
on the parameters $\alpha$, $M_0$, and $r_0$.
As predicted in Section~\ref{subsec:34},
we find that the dependence on the viscosity parameter $\alpha$
is slight, especially for $M_p > 3 \times 10^{-4}M_*$.
The dependence on the initial planet mass $M_0$ is also 
very weak when $M_p \gg M_0$.
The time evolution and the final value of the planet mass
is weakly dependent on $r_0$ for $M_p \gtrsim 10^{-2}M_*$.
This is because the ratio $D'/3\pi \nu$ is proportional to $r_p^{1/4}$.
Hence, we can say that the dependence of the final planet mass
on these parameters is weak.
As a result of the independence of $\alpha$,
the final planet mass is also approximately independent
of the initial disk radius, $R_d(t_0)$,
because $R_d(t_0)$ depends on $\alpha$ (see Equations~(\ref{rdss}) 
and (\ref{viscosity})).

Figure~\ref{fig:4}
shows the final planet mass as a function of the initial disk mass 
$M_d(t_0)$ in the case of $\dot{M}\sub{w}=10^{-9} M_* /\mbox{yr}$.
If all of the gas supplied from the outer disk perfectly accretes onto the planet,
the final planet mass is $M\sub{sup}+M_0$,
where $M\sub{sup}$ is given by Equation~(\ref{msup})\footnote{Mordasini 
et al.~(2009) also assumed perfect accretion
in their population synthesis calculations.}.
This approximate final planet mass is also plotted.

Both lines agree well with each other for $M_p \lesssim 10^{-3}M_*$
because the assumption of perfect accretion onto the planet
is valid for such less massive planets (see Section~\ref{subsec:34}).
For $M_p \gtrsim 10^{-2}M_*$, on the other hand,
a major part of the gas supplied from the outer disk does not 
accrete onto the planet but flows to the innermost disk. That is, 
the accretion onto the planet is imperfect in this case.

Note that, for a much less massive disk with 
$M_d(t_0) < 2t_0 \dot{M}\sub{w}$ ($=4\times 10^{-3} M_*$ in this case), 
the planet cannot grow through gas accretion because of early gas dissipation 
at the planet-forming region due to photoevaporation. Then, 
such a less massive disk produces no giant planet.
For $M_d(t_0) = 0.01M_*$, the final planet mass 
(or the supplied mass) is only approximately 10\% of $M_d(t_0)$. 
In this case, the disk mass at $t\sub{end}$ is 
approximately 70\% of $M_d(t_0)$ and 20\% of $M_d(t_0)$ is dissipated by the 
photoevaporation.

Figure~\ref{fig:5} shows the dependence of the final planet 
mass on the initial disk mass but for various mass loss rates 
$\dot{M}\sub{w}$.
From this result, we are able to determine how massive a disk is 
required at $t_0$ for each final planet mass.
The required disk mass increases with
the mass loss rate $\dot{M}\sub{w}$ for a given final planet mass.
In these mass loss rates,
the final planet mass is always less than or equal to 20\% of 
$M_d(t_0)$. 
For Jupiter-mass planets ($M_p\simeq 10^{-3}M_*$),
the ratios $M_{p,f}/M_d(t_0)$ are only 10, 5, and 2\%, 
with $\dot{M}\sub{w}=10^{-9}$, $3\times 10^{-9}$, and $10^{-8}M_*/$yr,
respectively. In Paper 2, we expected that the final planet mass 
should be comparable to the disk mass at $t_0$. This expectation 
was inaccurate because Paper 2 did not include photoevaporation.
Although Paper 2 suggested a much less massive disk than the MMSN disk
for the formation of Jupiter, our new model requires the mass 
(or the surface density) of the MMSN disk for Jupiter because of mass 
loss due to photoevaporation.
Even for massive disks with $M_d(t_0) \sim 0.1M_*$, 
the ratio, $M_{p,f}/M_d(t_0)$, does not increases much
because of the imperfect accretion onto massive planets,
as shown in Figure~\ref{fig:4}.
Thus, $M_{p,f}/M_d(t_0)$ is always kept small
by the mass loss due to photoevaporation 
and imperfect accretion onto the planet.

Figure~\ref{fig:6} shows the disk dissipation time, $t\sub{end}$,
for the planet-forming region due to photoevaporation.
The disk dissipation time is shortened for a higher mass loss rate
$\dot{M}\sub{w}$.
For a much lower mass loss rate than
$10^{-9} M_*/$yr, the disk lifetime is too long 
($t\sub{end} \gg 2\times 10^7$ yr).
Such a weak photoevaporation with a rate of $\leq 3\times 10^{-10} M_*/$yr
would be inconsistent with the observed disk lifetime.
Moreover, too strong photoevaporation with a rate of 
$\geq 3\times 10^{-8} M_*/$yr
would be inconsistent. Such strong photoevaporation would dissipate
the disk before the planetary cores grow to the critical core mass.
Thus, we can say that the mass loss rates in Figure~\ref{fig:5}
cover the entire reasonable range.
This reasonable range of $\dot{M}\sub{w}=10^{-9}$-$10^{-8} M_*/$yr
is consistent with previous estimations
(e.g., Armitage et al.~2003; Mordasini et al.~2009).
\begin{figure}[t]
\begin{center}
%
\includegraphics[width=8.3cm]{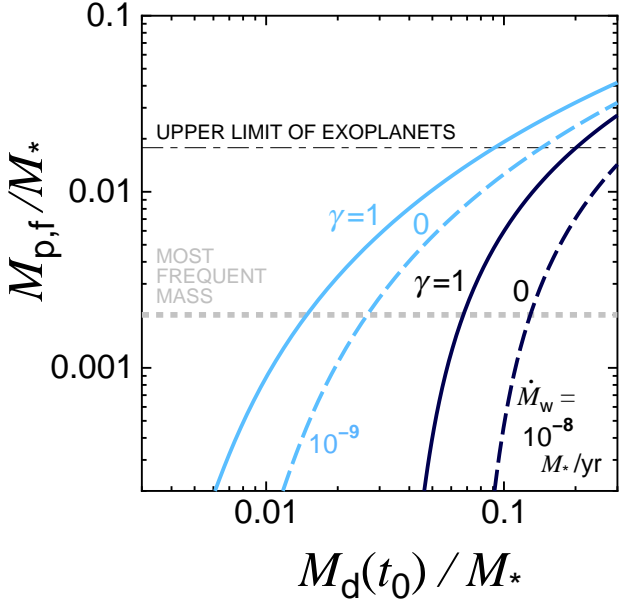}
%
\caption{Dependence of the final planet mass on the disk 
surface density gradient.}
\label{fig:8}
\end{center}
\end{figure}

Although, thus far, we have assumed $\gamma=1$ (i.e. $\nu \propto r$),
we also examine the $\gamma$-dependence of the final planet mass.
The index $\gamma$ also determines the radial surface density 
distribution as $\Sigma \propto r^{-\gamma}$.
In Figure~\ref{fig:8}, we also plot the final planet masses
for $\gamma=0$ (i.e., for a flat surface density distribution)
as an extreme case.
The flat surface density case produces
planets with relatively low masses as compared with the case of $\gamma=1$.
This is because the disk dissipation is earlier in $\gamma=0$,
as shown in Equation~(\ref{tend2}).
Since the case of $\gamma=0$ would be extreme,
the $\gamma$-dependence of the final planet mass would not be so
significant.

\subsection{Which Disks Produce Giant Exoplanets?} 
\label{subsec:42}
With our planet formation model, we can connect the data of 
giant exoplanets to observed disk masses.
In Figure~\ref{fig:5}, we also plot two reference planet masses,
the upper limit mass ($0.018M_*$) and most frequent mass ($\simeq 0.002M_*$)
of giant exoplanets.
These masses are obtained from data of exoplanets
observed by the radial-velocity methods (in Figure~\ref{fig:1}).

The required disk mass for the upper limit of exoplanets
is 0.1-0.2 $M_*$ for 
the reasonable mass loss rates of 
$\dot{M}\sub{w}=10^{-9}$-$10^{-8} M_*/$ yr.
The mass of the most massive observed disk
is also $\simeq 0.1 M_*$.
Thus, our model succeeds in reproducing the most massive exoplanets
within the observed disk masses and the reasonable disk mass loss rates.
It is also expected that exoplanets with the most frequent mass 
might be produced in disks with the most frequent disk mass,
which is $\sim 0.03M_*$ 
for a relatively old star-forming region of $\sim 10^6$ yr 
(Andrews et al.~2010).
This indicates that the disk mass loss rate of 
a few $10^{-9} M_*/$yr
is plausible for reproducing the most frequent exoplanets
in Figure~\ref{fig:5}.

\begin{figure}[t]
\begin{center}
%
%
\includegraphics[width=8.3cm]{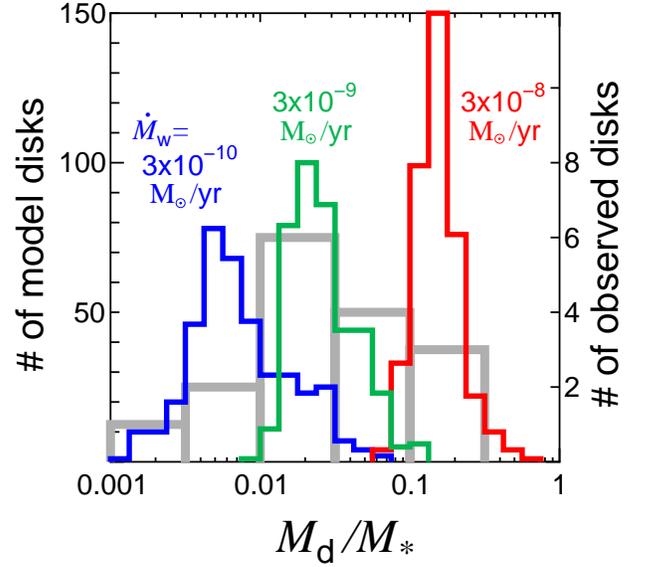}
%
\caption{Mass distributions of model disks 
forming exoplanets observed by the radial velocity survey.
The disk masses at $t_0$ are estimated using our simple model
for $\dot{M}\sub{w} = 3\times 10^{-10}$(blue), $3\times 10^{-9}$(green), 
and $3\times 10^{-8}M_\odot/$yr (red).
For other parameter settings, see the text.
The disk mass is normalized by the mass of the central star.
For comparison, we also plot the mass distribution of disks
observed in the Ophiuchus star-forming region (gray)
given by Andrews et al.~(2010).}
\label{fig:9}
\end{center}
\end{figure}

We also estimate the masses of the disks  
forming each exoplanet in Figure~\ref{fig:1} 
for a given mass loss rate $\dot{M}\sub{w}$,
using our simple model.
As for data of exoplanets, we use 
the planet mass $M_p \sin i$, the orbital radius (semi-major axis) 
$r_p$, and the mass of the central star $M_*$ obtained from 
the radial velocity survey data in http://exoplanets.org.
We simply put sin $i =$ 1.

We set $M_0$, $\alpha$, and $\gamma$ to be nominal values.
The estimated disk masses depend primarily on the masses of exoplanets
and the mass loss rate $\dot{M}\sub{w}$.
As shown in Figure~\ref{fig:3}, 
the orbital radii of exoplanets and
the parameters of $M_0$ and $\alpha$
do not greatly affect the estimation of disk mass.
Planets less massive than $M_0$ are excluded from our estimations.
We consider exoplanets with $r_p > 0.1$ AU only.
Disk mass estimation is performed for three mass loss rates of 
$\dot{M}\sub{w} = 3\times 10^{-10}$, $3\times 10^{-9}$, 
and $3\times 10^{-8}M_\odot/$yr.
The starting time of runaway gas accretion onto planets, $t_0$,
is set to be $2 \times 10^6$ yr.

Figure~\ref{fig:9}
shows the distributions of the estimated masses of 
the disks forming each exoplanet.
The disk mass is normalized by the mass of the central star.
For comparison, we also plot the mass distribution of disks
observed in the Ophiuchus star-forming region obtained by Andrews et al.~(2010).
Note that the disk masses are normalized by the masses of each central star.
The distribution obtained for $\dot{M}\sub{w} = 3\times 10^{-9}M_\odot/$yr
has almost the same peak as the observed disks.
In this case, the maximum disk masses in our model (0.13$M_*$) and the 
observation (0.24$M_*$) are also close to each other.
These results are consistent with 
%
Figure~\ref{fig:5}.
The obtained distribution has no disks with 
$M_d < 0.9 \times 10^{-3}M_*$ for 
$\dot{M}\sub{w} = 3\times 10^{-9}M_\odot/$yr.
This is because giant planets are not formed in 
less massive disks 
with $M_d(t_0) < 2t_0 \dot{M}\sub{w}$ ($=0.012 M_\odot$ in this case)
due to early disk dissipation (see Section~\ref{subsec:33}).
Only planets of Neptune size or less
are formed in such disks.

For $\dot{M}\sub{w} = 3\times 10^{-8}M_\odot/$yr, 90\% of disks have 
masses greater than 0.1$M_*$. Some are estimated as $M_d > 0.5M_*$.
Thus it is difficult to explain the origin of giant exoplanets
with the observed disks under the relatively high mass loss rate of 
$3\times 10^{-8}M_\odot/$yr. For a low mass loss rate of 
$3\times 10^{-10}M_\odot/$yr, on the other hand,
the obtained mass distribution has a peak at $5\times 10^{-3}M_*$,
which is inconsistent with the observed disks.
This low mass loss rate also causes a very long disk lifetime
of $3\times 10^7$yr for massive disks with 0.07$M_*$.
In addition, from Figure~\ref{fig:9}, therefore, we can conclude that
the mass loss rate of $3\times 10^{-9}M_\odot/$yr
is plausible to explain giant exoplanets
with realistic disks in our simple model.

\section{Summary and Discussion}

We developed a new model for giant planet formation by including 
the effect of photoevaporation and a new model for type II 
planetary migration proposed by Kanagawa et al.~(2018).
As an effect of photoevaporation, we included disk mass loss 
with a constant rate outside of the planet orbit. This mass loss
dissipates the disk gas at the planet-forming region and terminates 
planet growth. Our model can predict the final mass of a giant 
planet produced in a given disk. Our results are summarized as follows.

\begin{enumerate}
\item
Our simple model gives analytical and universal evolution tracks of 
planets growing via runaway gas accretion in the mass-orbit diagram 
(Equation~(\ref{evo_curve}) and Figure~\ref{fig:1}), which are completely 
independent of disk properties (i.e., the mass, radius,
temperature, or viscosity). 
Planets with a few Jupiter masses or less
suffer only a slight radial migration
at the runaway gas accretion stage because of their rapid growth.
Even the massive exoplanet with $\sim 20M_J$ at 3 AU
is formed from solid cores initially located within 20 AU.
Giant exoplanets crowded around 2 AU can be explained if a large 
number of massive solid cores are formed from 1.5 AU to 4 AU of 
protoplanetary disks. Such massive planetary embryos may be 
formed just outside of the snow line, which is located at 1 to 3 AU.
 
\item
We examined the time evolution and the final mass of a planet,
using a simple disk model including the effects of photoevaporation and 
gas accretion onto the planet. The final planet mass depends primarily on 
two parameters only (Figure~\ref{fig:5}). One is the product of the 
starting time of accretion onto the planet, $t_0$, and the mass loss rate 
due to photoevaporation, and the other is the initial disk mass at $t_0$.
The final planet mass depends only slightly on the initial disk radius, 
the viscosity, the initial planet mass, and the initial orbital 
radius (Figure~\ref{fig:3}). 
The mass loss rate due to photoevaporation is a parameter, but is 
constrained in the range of $10^{-9}$-$10^{-8} M_*/$yr by the lifetime 
of the observed disks (Figure~\ref{fig:6}).

\item
Giant planets grow through the accretion of the disk gas supplied 
from the outer disk. Planets of Jupiter-mass or less can capture 
the supplied disk gas almost perfectly. The final masses of such small 
planets are given by the total supplied mass $M\sub{sup}$ of 
Equation~(\ref{msup}). For massive planets with several Jupiter 
masses or larger, a major part of the gas passes by the planet and 
flows into the innermost disk (Figure~\ref{fig:4}).

\item
The ratio of the final planet mass to the initial disk mass at $t_0$
is always $\lesssim 0.1$, because of photoevaporation
and imperfect accretion onto the planet (Figure~\ref{fig:5}).\vspace{1mm}

\item
With our formation model, we can connect the
data of giant exoplanets to the observed disk masses.
The most massive exoplanet ($\simeq 20M_J$) is born in
the most massive T~Tauri disk with $M_d \sim 0.1M_*$
for the reasonable range of mass loss rate due to photoevaporation.
Our model also succeeds in explaining the most frequent mass 
of giant exoplanets ($\sim 2M_J$) with the most common disk mass with 
$\sim 0.02M_*$ for a disk mass loss rate of $3\times 10^{-9}M_\odot/$yr
(Figures~\ref{fig:5} and \ref{fig:9}).
Thus our model successfully explains properties in the mass distribution 
of giant exoplanets with the observed mass distribution of protoplanetary 
disks for a reasonable range of the mass loss rate due to photoevaporation.

\end{enumerate}

In our simple model, we focused on the formation of a single giant 
planet in each protoplanetary disk and did not consider any 
interaction between multiple planets.
Interactions between multiple planets, however, can be important to 
explain observed giant exoplanets. Due to their gravitational 
interactions, multiple gas giant systems can be orbitally 
unstable. The orbital instability of such multiple systems
often produces giant planets in eccentric orbits and ejects some planets 
from the system at the same time (jumping Jupiters model; 
e.g., Rasio \& Ford 1996; Weidenschilling \& Marzari 1996; Lin \& Ida 1997).
Planet-planet scattering followed by tidal circularization can 
form hot Jupiters with small orbital radii (e.g., Rasio \& Ford 1996; 
Nagasawa et al.~2008; Winn et al.~2010). Since our universal evolution tracks 
indicate insufficient type II migration for planets of Jupiter-mass or 
less because of their rapid growth via runaway gas accreion
(Figure~\ref{fig:1}), 
the scenario of planet-planet scattering would be plausible for hot Jupiters.
The type I migration of solid planetary cores before
the runaway gas accretion stage can be an alternative possible origin 
of hot Jupiters since their growth time is expected to be much longer
than that at the runaway gas accretion stage.

Indirect interaction through gas accretion onto planets would 
also be important. We consider multiple giant planets growing by 
gas accretion. If the outermost giant planet has a Jupiter 
mass or less, the gas accretion onto the outer planet is almost perfect 
(Figure~\ref{fig:4}) and the mass supply to inner planets is 
significantly reduced. Thus, only the outermost giant planet can grow with 
gas accretion in this system. When the outermost giant planet has 
grown to several Jupiter masses or larger, however, its gas accretion 
becomes imperfect and the inner planets can also grow. Jupiter and Saturn 
are also expected to have been influenced by such an indirect interaction 
in their growth stages. Once Saturn starts runaway gas 
accretion, Jupiter, which is located inside Saturn's orbit, 
cannot grow further due to the perfect accretion onto Saturn. 
Thus, Saturn's runaway gas accretion should start just after 
Jupiter had grown to its present mass.
Such a formation scenario for Jupiter and Saturn is suggested 
by our simple model. Both direct and indirect interactions should be 
included in future formation models for multiple systems of giant planets.

The empirical formula of the gas accretion rate onto a planet
used in our model might need to be improved.
Our empirical formula of the mass  accretion rate has been confirmed by the 
hydro-dynamic simulations by D'Angelo et al.~(2003) and 
Machida et al.~(2010), but only for Jupiter-mass planets or smaller.
Only a few hydro-dynamic simulations have been performed for gas accretion onto a 
giant planet much heavier than Jupiter.
Kley \& Dirksen (2006) showed through their hydro-dynamical simulations that 
massive giant planets with masses $\ge 3M_J$ strongly excite
an eccentric motion of gas at the edge of the planetary gap.
Owing to the eccentric motion, the gas accretion rate onto the planet is 
greatly enhanced in their simulations with the planet masses $\ge 5M_J$.
On the other hand, Bodenheimer et al.~(2013) obtained much lower 
gas accretion rates than in our formula in their hydro-dynamical 
simulations with $\ge 3M_J$.
Further extensive hydro-dynamical simulations 
on gas accretion onto high-mass giant planets
should be performed in order to fix the accretion rate accurately.\vspace{5mm}

%
The authors would like to thank anonymous referees
for valuable comments. The authors also thank Shigeru Ida and Kazuhiro Kanagawa 
for providing detailed comments.
The present study was supported by JSPS KAKENHI grant numbers
17H01103, 18H05438, 19K03941, and 15H02065.

\end{document}